\documentclass[twocolumn,showpacs,preprintnumbers,amsmath,amssymb,prl]{revtex4}

\usepackage{graphicx}
\usepackage{dcolumn}
\usepackage{bm}
\usepackage{amsmath}
\def\degC{\kern-.2em\r{}\kern-.3em C}

\begin{document}

\preprint{******}

\title{Ultraslow Propagation of Squeezed Vacuum Pulses with \\Electromagnetically Induced Transparency}

\author{Daisuke Akamatsu$^{1}$, Yoshihiko Yokoi$^{1}$, Manabu Arikawa$^{1}$, \\ Satoshi Nagatsuka$^{1}$, Takahito Tanimura$^{1}$, Akira Furusawa$^{2}$, and Mikio Kozuma$^{1,3}$}

\affiliation{%
$^{1}$Department of Physics, Tokyo Institute of Technology, 2-12-1 Okayama, Meguro-ku, Tokyo 152-8550, Japan}

\affiliation{%
$^{2}$Department of Applied Physics, School of Engineering, The University of Tokyo, 7-3-1 Hongo, Bunkyo-ku, Tokyo 113-8656, Japan}
\affiliation{%
$^{3}$PRESTO, CREST, Japan Science and Technology Agency, 1-9-9 Yaesu, Chuo-ku, Tokyo 103-0028, Japan}%

\date{\today}

\begin{abstract}
We have succeeded in observing ultraslow propagation of squeezed vacuum pulses with electromagnetically induced transparency. Squeezed vacuum pulses (probe lights) were incident on a laser cooled $^{87}$Rb gas together with an intense coherent light (control light). A homodyne method sensitive to the vacuum state was employed for detecting the probe pulse passing through the gas. A delay of 3.1 $\mu$s was observed for the probe pulse having a temporal width of 10 $\mu$s. 
\end{abstract}

\maketitle
Electromagnetically induced transparency (EIT) is used to modify the absorption coefficient and refractive index of a medium for a probe light \cite{EIT}. There is a steep dispersion within the transparency window, so that the speed of a probe light pulse is significantly reduced. The pulse is spatially compressed as a consequence of this ultraslow propagation, which enables us to store photonic information in the medium \cite{storage of light, dark state polariton}. 

Recently, several groups have independently succeeded in storing and retrieving a single-photon state by collective atomic excitation \cite{Kuzmich,Lukin}. These experiments have demonstrated nonclassical characteristics of the retrieved light field, such as photon antibunching and the violation of classical inequalities. Such nonclassical features are not sensitive to linear optical loss, and the experiments were performed using a photon counting method. In order to fully determine the state of the field, we have to utilize an optical homodyne method, which is sensitive to the vacuum. While B. Julsgaard \textit{et al.} have demonstrated quantum memory of light using a homodyne method \cite{Polzik-QM}, they have thus far only reported an experiment with a coherent state of light. 

While EIT was observed using a squeezed vacuum in our previous study \cite{Akamatsu}, storage and retrieval of a squeezed vacuum could not be realized, mainly due to poor squeezing of the light source as well as the transparency window being too broad. In the present Letter, we report the successful observation of the ultraslow propagation of a squeezed vacuum pulse, which is an important step toward the demonstration of a genuine quantum memory for non-classical light states. The storage of the squeezed vacuum corresponds to that of quantum entanglement and enables us to squeeze atomic spins deterministically, which is useful for quantum noise-limited metrology. Performing a single photon count for the field retrieved partially from the squeezed atoms, a highly nonclassical atomic state will be generated \cite{Polzik-cat}.

\begin{figure}
    \includegraphics[width=\linewidth]{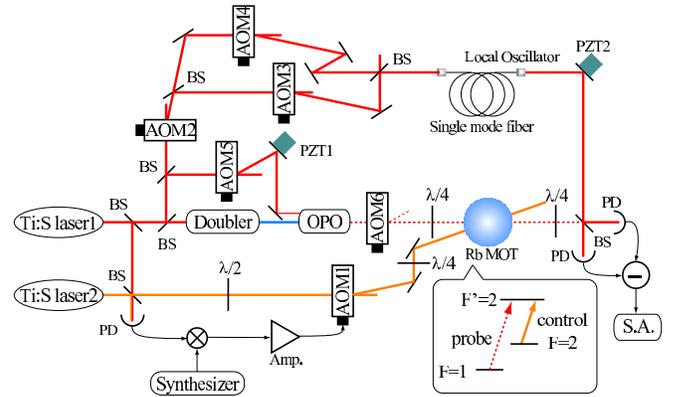}
  \caption{(color online) Schematic diagram of the experimental setup. BS: beam splitter, AOM: acousto-optic modulator, PD: photodetector, SA: spectrum analyzer, Amp.: RF amplifier. AOM5 consists of two AOMs to cancel the frequency shift due to the diffraction. AOM6 also consists of two AOMs, where, rather than the 1st-order beam, the 0th-order beam was used.}
    \label{fig:1.eps}
\end{figure}

Our experimental setup is shown schematically in Fig. 1. A Ti:sapphire laser (Ti:S laser 1) was tuned to the D$_1$ line (5$^2$S$_{1/2}$, F=1 $\to$ 5$^2$P$_{1/2}$, F'=2), which corresponds to a probe transition. The beam from the other Ti:sapphire laser (Ti:S laser 2) was diffracted by an acousto-optical modulator (AOM) 1 and was used for the control field. The frequency of the control light was stabilized using a feed-forward method \cite{Kourogi}, and the control light was able to be scanned around the F=2 $\to$ F'=2 transition by tuning the frequency of a synthesizer. We employed a laser-cooled atomic ensemble of $^{87}$Rb as an EIT medium. One cycle of our experiment comprised a cold atom preparation period and a measurement period. The preparation period and the measurement period had durations of 8.7 ms and 1.3 ms, respectively. After 5.5 ms of the magneto-optical trapping stage in the preparation period, only the magnetic field was turned off. After the eddy current ceased ($\sim$ 3 ms), both the cooling and repumping lights were turned off, and a pump light, which was tuned to F=2 $\to$ F'=2 transition, was incident on the gas for 100 $\mu$s to prepare the cold atoms in the F=1 state, the optical depth of which was $\sim$ 4.

A weak coherent probe light was used to observe the frequency width of the EIT window. Note that, in the later experiment, we used a squeezed vacuum for the probe light, where the squeezed vacuum was generated using a sub-threshold optical parametric oscillator (OPO) with a periodically poled KTiOPO$_{4}$ crystal \cite{Tanimura}. We injected a light in a coherent state into the OPO cavity in the absence of a second-harmonic light from a doubler and used the output as the probe light. The procedure described above enabled us to employ a coherent probe light with a spatial mode that was identical to that of the squeezed vacuum. The cavity length was actively stabilized so that the probe frequency was equal to the resonant frequency of the cavity. The probe light ($<$1 pW) from the OPO cavity and the control light (100 $\mathrm{\mu}$W) were incident on the gas with a crossing angle of 2.5$^\circ$. The radii of the probe and the control lights were 150 $\mathrm{\mu}$m and 550 $\mathrm{\mu}$m, respectively. Both the probe and the control lights were circularly polarized in the same direction. During the measurement period, the probe light was incident on the atomic gas and its transmitted intensity was monitored using an avalanche photodiode (not shown in Fig. 1). Figure 2(a) represents a typical transmission spectrum for the probe light obtained by scanning the frequency of the control light, where the medium was almost transparent around two-photon resonance. 
\begin{figure}
    \includegraphics[width=.7\linewidth]{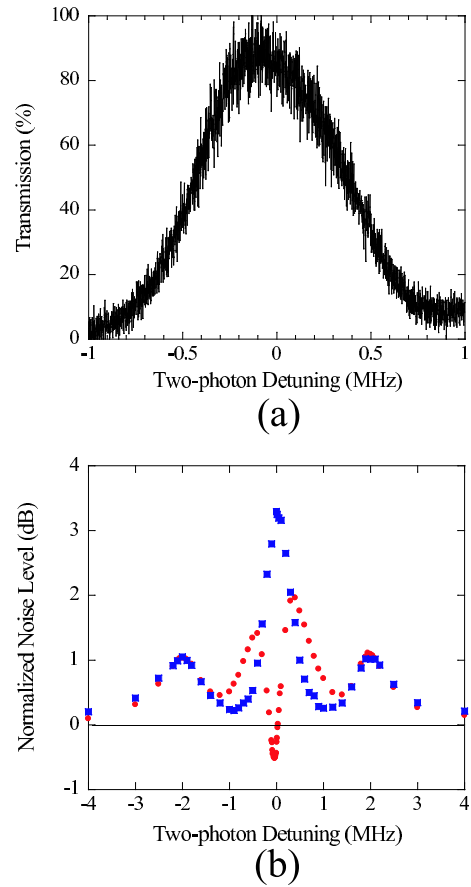}
  \caption{(color online)(a) Dependence of the transmission of the coherent probe light on two-photon detuning. (b) Quadrature noise of the probe light in the squeezed state that passed through the EIT medium. Circles (Squares) indicate the quadrature noises when the relative phase between the LO and the squeezed vacuum was set to $\theta=0$ ($\pi/2$).}
    \label{fig:2.eps}
\end{figure}

In order to observe ultraslow propagation of a squeezed vacuum pulse with EIT, it is necessary to measure the quadrature noise of the frequency components within the EIT window. However, it is difficult to observe the quadrature squeezing for such a low-frequency region by the conventional homodyne method using a spectrum analyzer, which is due to the low-frequency noise of a spectrum analyzer. While several schemes are available \cite{Lvovsky, Polzik-cat}, we employed the homodyne method with a bichromatic local oscillator (LO) to observe quadrature squeezing around the carrier frequency. The frequency of the LO of the conventional homodyne method is the same as the carrier frequency of the squeezed vacuum $\nu_0$. The noise power measured using a spectrum analyzer having a center frequency of $\epsilon$ in the zero span mode is given by\cite{Yurke}
\begin{eqnarray}
S_\epsilon(\theta)=\mathrm{Tr}[\rho[\hat{X}(\epsilon,\theta)\hat{X}^\dag(\epsilon,\theta)+\hat{X}^\dag(\epsilon,\theta)\hat{X}(\epsilon,\theta)]].
\label{eq:nondegenerate}
\end{eqnarray}
Here, $\rho$ represents the density operator of the squeezed vacuum and $\theta$ represents the relative phase between the squeezed vacuum and the LO. The quadrature operator is defined as $\hat{X}(\epsilon,\theta)=\hat{a}_{\nu_0+\epsilon}e^{-i\theta}+\hat{a}^\dag_{\nu_0-\epsilon}e^{i\theta}$. However, when the bichromatic LO having the frequency components of $\nu_0\pm\epsilon$ is employed, the noise measured by the spectrum analyzer with the center frequency of $\epsilon$ is given by
\begin{align}
S(\theta)&=\mathrm{Tr}[\rho[(\hat{X}(0,\theta)\hat{X}^\dag(0,\theta)+\hat{X}^\dag(0,\theta)\hat{X}(0,\theta))/2\nonumber\\
&+(\hat{X}(2\epsilon,\theta)\hat{X}^\dag(2\epsilon,\theta)+\hat{X}^\dag(2\epsilon,\theta)\hat{X}(2\epsilon,\theta))/2]]. \label{eq:degenerate}
\end{align}
Comparing Eq. (\ref{eq:nondegenerate}) and Eq. (\ref{eq:degenerate}), the following simple formula can be obtained, $S(\theta)=S_0(\theta)/2+S_{2\epsilon}(\theta)/2$. The first term represents the single-mode quadrature noise at $\nu_0$, and the second term represents the two-mode quadrature noise consisting of $\nu_0 \pm 2 \epsilon$. Namely, both the carrier and sideband frequency components contribute to the measured noise. When the medium has sufficient optical thickness and the width of the EIT window is narrower than $2\epsilon$, the sideband frequency components are completely absorbed and thus the noise amplification and deamplification are caused by only the carrier frequency component. Note that in such a case, the second term in Eq. (\ref{eq:degenerate}) approaches 1/2 (the vacuum noise) and the observable squeezing level is limited to $-3$ dB.

To check the validity of the method described above, we measured the squeezing using both a monochromatic LO and a bichromatic LO. We cut off the weak coherent light and generated a squeezed vacuum by injecting 50 mW of the pump light (397.5 nm) from the frequency doubler into the OPO cavity. We employed three AOMs (AOM2-4) to create the monochromatic and bichromatic LOs. The monochromatic LO was produced by turning off AOM4 and driving AOM2 and AOM3 with the same RF frequency, while the bichromatic LO was obtained by driving AOM2, 3, and 4 with RF frequencies of 80 MHz, 79 MHz, and 81 MHz, respectively.
\begin{figure}
    \includegraphics[width=.7\linewidth]{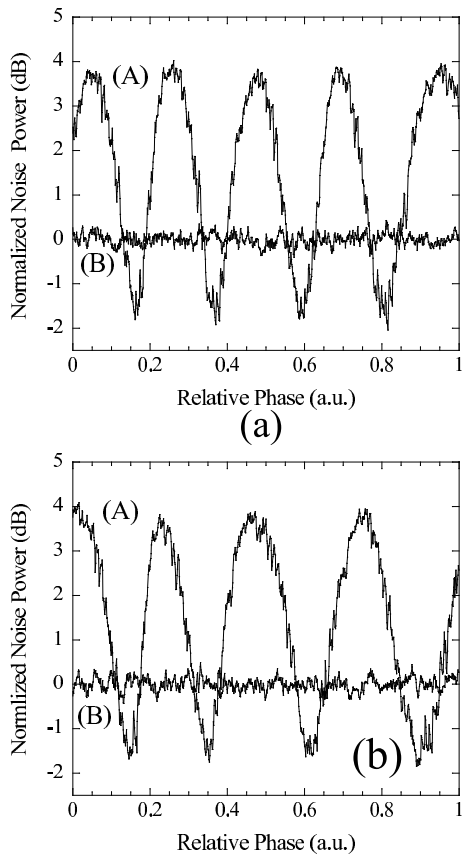}
  \caption{(A) Balanced homodyne signals of the probe light with (a) a monochromatic LO and (b) a bichromatic LO. (B) Shot noise level. We normalized the quadrature noise levels using the shot noise.}
    \label{fig:3.eps}
\end{figure}

We measured the two-mode quadrature noise consisting of the $\nu_0 \pm $1 MHz frequency components using the monochromatic LO (Fig. 3(a)). To determine the shot noise level, we measured the noise dependence on the intensity of LO and confirmed that the noise level agreed well with a theoretical calculation. The squeezing level of $-$1.60$\pm$0.12 dB and the anti-squeezing level of 3.71$\pm$0.12 dB were observed with the spectrum analyzer in the zero-span mode at $\epsilon$=1 MHz. The resolution bandwidth was 100 kHz and the video bandwidth was 30 Hz. The 3-dB bandwidth of our homodyne detector was 2 MHz. Figure 3(b) shows the experimental results obtained using the bichromatic LO, where the squeezing level of $-$1.53$\pm$0.20 dB and the anti-squeezing level of 3.66$\pm$0.21 dB were observed. The LO intensity at $\nu_0\pm1$ MHz was 1.5 mW each, and the total intensity of LO was monitored and stabilized so that the fluctuation was less than 1\%. Note that the linewidth of the OPO cavity (10 MHz) was much larger than 2 MHz (2$\epsilon$), and thus $S_0(\theta)$ and $S_{2\epsilon}(\theta)$ should have a squeezing level comparable to that obtained in Fig. 3(a). Therefore, there is no discrepancy between the experimental results and the prediction by Eq. (\ref{eq:degenerate}). 

In order to carry out the EIT experiments with a squeezed vacuum, the relative phase between the local oscillator light and the squeezed vacuum has to be stabilized during the measurement period. For this purpose, a weak coherent beam (lock light) was injected to the OPO cavity, and the output was monitored by a photodetector (not shown in Fig. 1). The relative phase between the lock light and the second harmonic light was locked using PZT1 so that the maximum amplification (deamplification) was obtained. We locked a relative phase between the LO and the lock light by controlling the PZT2 based on the signal of the homodyne detector. This procedure enabled us to lock the relative phase between the LO and the squeezed vacuum at $\theta =0 (\pi/2)$ \cite{relative phase lock}. After the preparation period, the feedback voltage driving a PZT was maintained, and the weak coherent light was turned off with AOM5. Eventually, the relative phase between the LO and the squeezed vacuum was maintained during the measurement period \cite{Takei}.

The quadrature noises of the squeezed vacuum that passed through the cold atoms with the control light (100 $\mu$W) were monitored using the bichromatic homodyne method. Figure 2(b) indicates the dependence of the quadrature noise on the two-photon detuning, where the circles (squares) were obtained when the relative phase was set to $\theta=0$ ($\pi$/2). Each data was averaged over $\sim$100,000 times and both the resolution and video bandwidths of the spectrum analyzer were set to 100 kHz. When the relative phase was set to $\theta=0$, the squeezing level of 0.44 $\pm$ 0.09 dB was detected at the two-photon resonance and the squeezing level decreased with increased detuning, which reflects a property of the transparency window. At around 300 kHz of detuning, the quadrature noise exceeded the shot noise level because the EIT medium provided the additional phase to the probe light and changed the relative phase $\theta$. Another peak, which was concerned with two-mode quadrature noise $S_{2\epsilon}$ ($\epsilon=1$ MHz), appeared around $\pm$2 MHz. When the control light was detuned by $\pm$ 2 MHz, the frequency component corresponding to $\nu_0\pm2$ MHz passed through the EIT medium, whereas that at $\nu_0\mp2$ MHz was absorbed. Therefore, the quantum correlation between the two frequency modes was lost, and the thermal noise corresponding to one frequency component was simply observed by the homodyne detector. Note that the noise levels were identical for both $\theta=0$ and $\pi/2$.

\begin{figure}[tbp]
    \includegraphics[width=.6\linewidth]{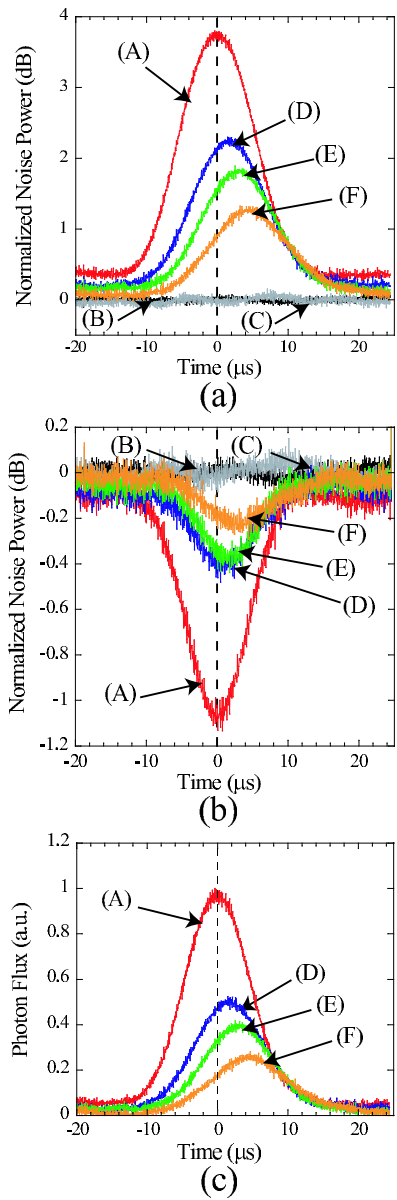}
  \caption{(color online) Time dependence of the measured noise of the probe pulse with the relative phase (a) $\theta=\pi/2$ and (b) $\theta=0$ and (c) that of the photon flux derived from data (a) and (b). Trace (A) shows the quadrature noises of the squeezed vacuum pulses in the absence of the control lights and the cold atoms. Trace (B) shows the quadrature noises of the squeezed vacuum pulses incident on the cold atoms without the control light. Trace (C) indicates the shot noises. Traces (D), (E), and (F) show the quadrature noises of the squeezed vacuum pulses incident on the cold atoms with control lights of intensities 200, 100, and 50 $\mu$W, respectively.}
    \label{fig:4.eps}
\end{figure}

In order to conduct ultraslow propagation of a squeezed vacuum pulse, we created a probe pulse having a temporal width of 10 $\mathrm{\mu}$s from the continuous-wave squeezed vacuum by using two AOMs in series (AOM6 in Fig. 1). Rather than the 1st-order diffracted light, we used the 0th-order (non-diffracted) light as the probe light. The diffraction efficiency of each AOM was 80\%, and thus the use of the 1st-order beam would have caused significant optical loss. Therefore, we used the 0th-order beam for the experiments. Figures 4(a) and (b) show the quadrature noise of the squeezed vacuum pulses with the relative phase of $\theta=\pi/2$ and 0, respectively. The signal was averaged over $\sim$100,000 measurements. Traces (A) and (B) in Fig. 4 show the quadrature noises of the squeezed vacuum pulses without and with the laser cooled gas in the absence of the control light, respectively. The optically dense medium absorbed the squeezed vacuum pulse, and thus trace (B) almost overlapped the shot noise (trace (C)). When the control lights were incident on the cold atoms, the transmitted squeezed vacuum pulse was delayed. The delay time increased as the intensity of the control light decreased (see traces (D), (E), and (F) in Fig. 4), which is a clear feature of slow propagation caused by EIT \cite{dark state polariton}. A maximum delay of 3.1 $\pm$ 0.11 $\mu$s was observed for the squeezed vacuum pulse with 50 $\mu$W of the control light (trace (F) in Fig. 4(b)). The photon flux of the probe light can be calculated by adding the quadrature noises of Figs. 4(a) and (b). Figure 4(c) clearly shows that the photon flux was also delayed.

In conclusion, we have successfully observed the squeezing of the probe light after passing through the sub-MHz EIT window with a bichromatic homodyne method. Ultraslow propagation of the squeezed vacuum pulses was demonstrated, where the delay time was increased by decreasing the intensity of the control light. A maximum delay of 3.1 $\mu$s for 10 $\mu$s pulses is sufficient for the storage of a squeezed vacuum pulse \cite{storage of light, Kuzmich, Lukin}. We are currently attempting to store and retrieve squeezed vacuum pulses with electromagnetically induced transparency. 

The authors would like to thank M. Kourogi, K. Imai, N. Takei, and K. Akiba for many stimulating discussions. One of the authors (D. A.) was supported in part by the Japan Society for the Promotion of Science. This study was supported in part by a Grant-in-Aid for Scientific Research (B) and by MEXT through the 21st Century COE Program ``Nanometer-Scale Quantum Physics'' at the Tokyo Institute of Technology

\end{document}